\documentclass[prb, twocolumn,superscriptaddress,preprintnumbers,a4paper,amsmath,amssymb,showpacs,floatfix]{revtex4}
\usepackage{graphicx}
\usepackage{color}\color[rgb]{0.000,0.000,0.000} 
\usepackage{amssymb} 
\usepackage{amsmath} 
\usepackage{multirow}

\begin{document}

\newcommand{\pb}{Pb$_{3}$Mn$_7$O$_{15}$}
\newcommand{\hex}{$P6_{3}/mcm$}
\newcommand{\six}{Ru$^{6+}_2$O$_9$}
\newcommand{\orth}{$Pnma$}
\newcommand{\te}{$_{t}$}
\newcommand{\ot}{$_{o}$}
\newcommand{\Tc}{T$_{C}$}
\newcommand{\Ts}{T$_{s}$}
\newcommand{\Tn}{$T_\mathrm{N}$}
\newcommand{\MuB}{$\mu_\mathrm{B}$}
\title{Charge and orbital order in frustrated \pb}
\author{Simon A. J. Kimber}
\email[Email of corresponding author:]{kimber@esrf.fr}
\affiliation{European Synchrotron Radiation Facility (ESRF), 6 rue Jules Horowitz, BP 220, 38043  Grenoble Cedex 9, France}

\date{\today}

\pacs{75.50-y, 75.25.Dk}
\begin{abstract}
The candidate magnetoelectric \pb \ has a structure consisting of 1/3 filled Kagom\'e layers linked by ribbons of edge-sharing octahedra in the stacking direction. Previous reports have indicated a complex hexagonal-orthorhombic structural transition upon cooling through $\sim$ 335 K, although its origins are uncertain. Here both structures are revisited using a combination of neutron and synchrotron X-ray diffraction data. Large shifts of oxygen positions are detected which show that the interlayer sites and those which occupy voids in the kagom\'e lattice are trivially charge ordered in both phases. The symmetry breaking is found to occur due to Mn$^{3+}$ orbital ordering on the ribbon sites and charge ordering of the sub-set of layer sites which make up a Kagom\'e network.  \end{abstract}
\maketitle
\section{\label{sec:level1}INTRODUCTION}
Transition metal oxides with spin, charge and orbital degeneracies often undergo symmetry breaking transitions on cooling \cite{attfield}. Especially complex examples occur when geometrical frustration hinders the long range ordering of one (or more) of these order parameters. This can lead to the emergence of technologically useful properties, for example the multiferroic behaviour found in the $RE$Mn$_{2}$O$_{5}$ ($RE$=Y-Lu) materials \cite{chapon}. Particularly rich behaviour is found in materials containing cations with lone pairs of electrons, such as Pb$^{2+}$ or Bi$^{3+}$. The Pb-Mn-O system is a case in point, with Pb$_{2}$MnO$_{4}$  predicted to show piezoelectricity and piezomagnetism \cite{kimber}, and dielectric anomalies \cite{dielectric} \ discovered in \pb. The latter is of special interest, as this material is mixed-valence and quasi-layered (Fig. 1), with 1/3 filled Kagom\'e layers connected by ribbons of edge-sharing MnO$_{6}$ octahedra and intercalated Pb$^{2+}$ cations. The exact crystal structure has been the subject of dispute for many years, and was originally reported as orthorhombic ($Cmc2_{1}$ or $Cmcm$) then revised to hexagonal  (\hex) by single crystal X-ray diffraction\cite{str1,str2,str3}. A breakthrough was made in 2009 by Rasch $et \ al$ \ who used powder synchrotron X-ray diffraction to confirm an orthorhombic metric and detect weak superstructure reflections that break the C-centring of the ortho-hexagonal $\sqrt{3}$.\textit{\textbf{a}} x \textit{\textbf{a}} x \textit{\textbf{c}} unit cell \cite{rasch}. Note that in the $Pnma$ \ setting the axes are interchanged such that the layers are stacked in the [100] direction, rather than in the [001] direction found in previous works. 
\begin{figure}[tb!]
\begin{center}
\includegraphics[scale=0.45]{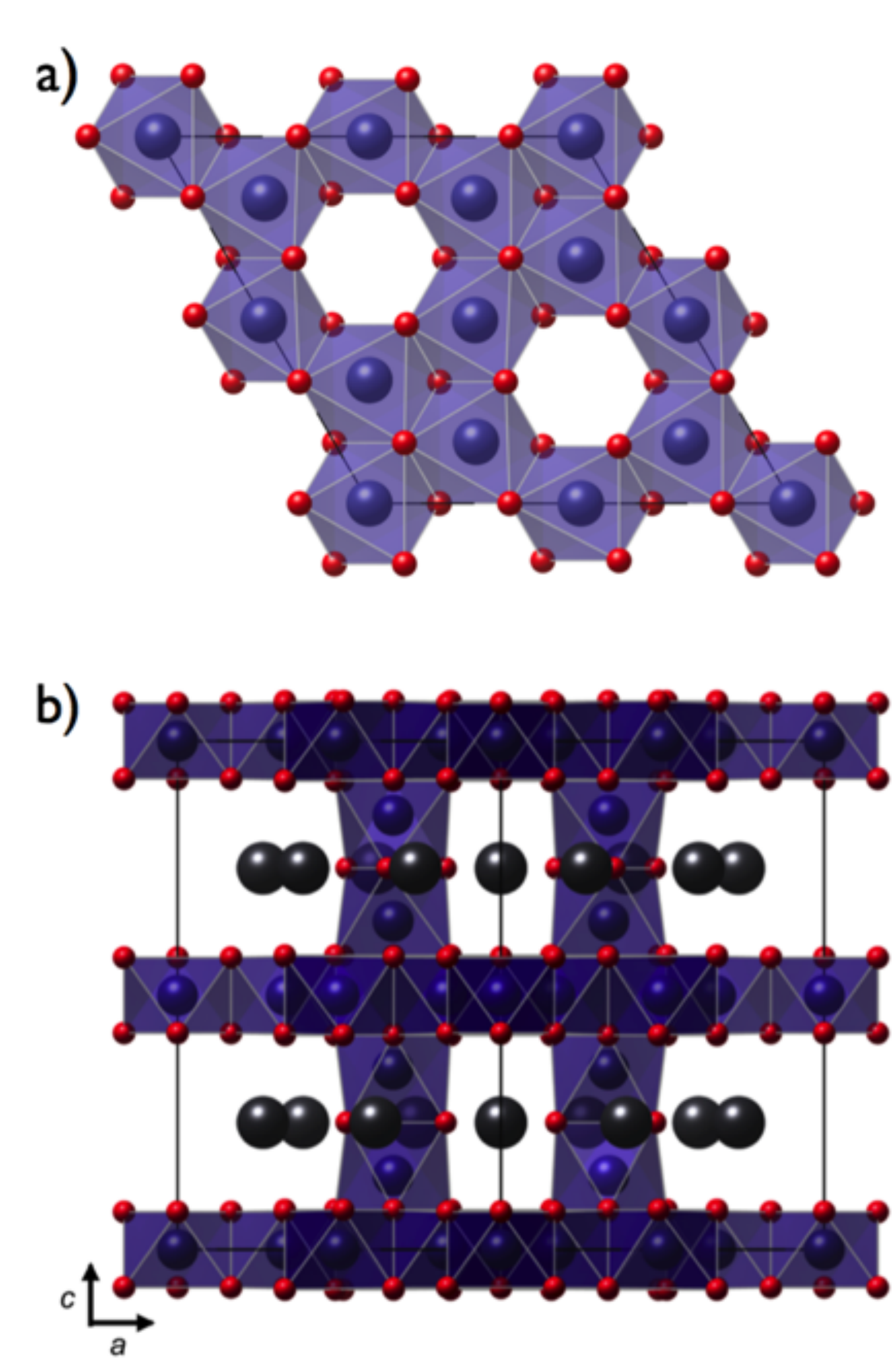}
\caption{(color online) a)  Edge sharing MnO$_{6}$ layers in  $P6_{3}/mcm$ \pb \ projected down [001]; b)  Structure of \pb \ projected down [110], MnO$_{6}$ octahedra and the associated oxygen atoms are shown, while Pb$^{2+}$ cations are represented by grey spheres.}
\label{Fig1}
\end{center}
\end{figure}
A structural phase transition to a hexagonal metric has additionally been shown to occur on warming above room temperature \cite{warm}.  However, while investigations of magnetic \cite{mag1,mag}, thermal \cite{hcap} \ and dielectric properties \cite{dielectric} \ have continued apace, there is as yet no consensus on the electronic factors which drive the \hex \ to \orth \ phase transition. This is because of the great complexity of the \orth \ structure, which contains 30 independent sites with 76 positional variables in a unit cell with V $\simeq$ 2300 \AA$^{3}$. Moreover, the detection of possible charge or orbital ordering relies upon accurate and precise measurements of oxygen displacements. This is extremely challenging by X-ray diffraction due to the presence of heavy elements. Meanwhile, it has recently been reported that the related material Pb$_{3}$Rh$_{7}$O$_{15}$  undergoes a resistive transition on cooling below 185 K, which is proposed to occur due to charge ordering \cite{rh,rh2}. This underscores the importance of determining the relationship between electronic degrees of freedom and crystal structure in this emerging class of materials.\\
Here a comprehensive re-investigation of the structure of \pb \ by combined Rietveld refinement \cite{multi} \ against synchrotron X-ray and multiple wavelength neutron powder diffraction is reported. This  complementary use of techniques is particularly suitable to charge and orbital ordering problems, where microtwinning, extinction, and multiple scattering, as well as subtle lattice distortions, hinder single crystal approaches \cite{jon,goff}. The results reported here confirm the space groups proposed at room temperature and above, however, shifts of oxygen positions of up to 0.15 \AA \ are found. The new structural models show that the phase transition is driven by charge and orbital order, providing a first insight into the electronic correlations in this complex material.
\section{\label{sec:level1}EXPERIMENTAL}
Polycrystalline \pb \ was synthesised from PbO (99.999\%, Aldrich) and Mn$_{2}$O$_{3}$ \ (99.999\%, Aldrich). Stoichiometric quantities of the reagents were intimately ground, pelleted and reacted under air for a total of four days at 830$\,^{\circ}\mathrm{C}$ with several intermediate regrinds. After each stage in the synthesis process the sample was cooled over several hours at a rate of 20$\,^{\circ}\mathrm{C}$/hour to 500$\,^{\circ}\mathrm{C}$ then the furnace was switched off. Synchrotron powder X- ray diffraction profiles were collected with the crystal analyser diffractometer ID31 at the ESRF, Grenoble, France. The wavelength of 0.45621(1) \AA  \ was calibrated using NIST standard silicon powder. The sample was held in a 0.2 mm borosilicate capillary and rotated to minimise preferred orientation. Data were collected for several hours at room temperature and more rapidly on cooling with a helium flow cryostat. Neutron powder diffraction profiles were collected with the E9 high resolution diffractometer at the Helmholtz-Zentrum Berlin, Germany. The sample was placed in a vanadium can of 6 mm diameter and data collected at room temperature and at 510 K with the aid of a cryofurnace. At 298 and 510 K, a wavelength of 2.806 \AA \ was selected with the (311) reflection of the germanium monochromator and data collected for
36 hours. Data were also collected for a further 24 hours at room temperature after selecting a wavelength of 1.79 \AA \ with the (511) monochromator reflection. The primary 10" collimator was inserted for all data sets to optimise resolution. Rietveld refinements of structural models against the X-ray and neutron data were performed using the GSAS suite of programs with the EXPGUI interface \cite{gsas,expgui}. The peak shapes for all data were modelled with a pseudo-Voigt peak profile function with the asymmetry correction of Finger $et\ al$ \cite{finger}. For the synchrotron X-ray data, for which the resolution reaches $\Delta$ d/d ~1x10$^{-5}$, an $hkl$ dependent Lorentzian broadening function was used. The wavelengths used for the neutron diffraction histograms were refined and the (calibrated) X-ray wavelength held fixed.  Crystal structures were visualised, and bond valence sums calculated, using the VESTA software package \cite{vesta}.
\begin{figure}[tb!]
\begin{center}
\includegraphics[scale=0.25]{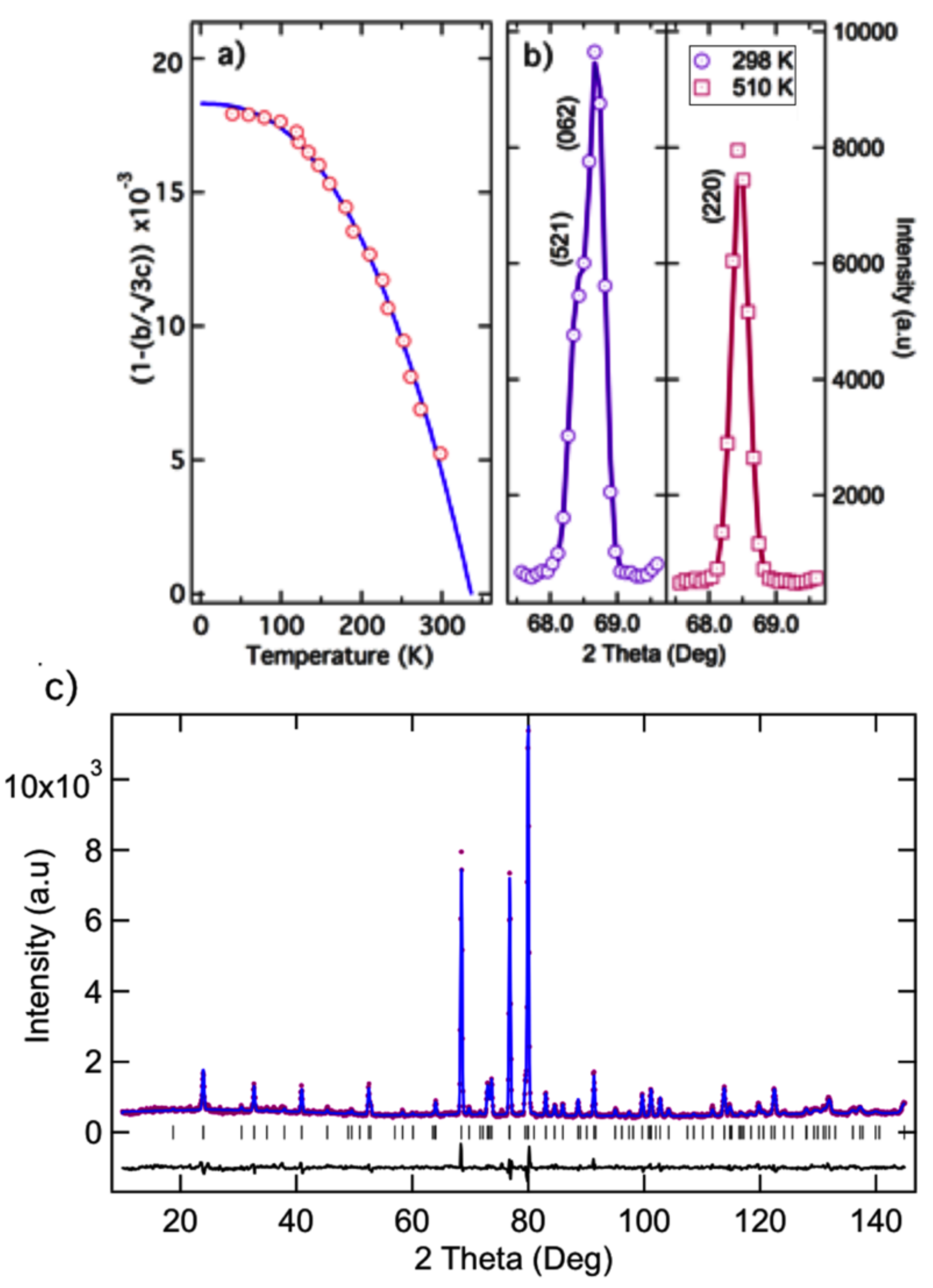}
\caption{(color online) a) Temperature dependence of the \textit{\textbf{b}}/\textit{\textbf{c}} ratio for \pb \ from neutron powder diffraction and synchrotron X-ray powder diffraction. The line is a power law fit used to estimate the transition to a metrically hexagonal cell. b) Neutron powder diffraction data at 298 and 510 K showing the (220)$_{hex}$ peak for \pb, and the orthorhombic splitting at 298 K. c) Observed, calculated and difference plots for the Rietveld fit of the \hex \ structure to the 2.8 \AA \ neutron powder diffraction profile of \pb \ at 510 K. Reflection positions are marked by vertical bars.}
\label{Fig1}
\end{center}
\end{figure}
\section{\label{sec:level1}RESULTS}
 All of the diffraction data collected at room temperature and below were consistent with the orthorhombic cell proposed by Rasch $et \ al$. However, the temperature dependence of the refined lattice parameters suggests that \pb \ is close to a structural phase transition. As shown in Fig. 2a, the \textit{\textbf{b}}/\textit{\textbf{c}} ratio approaches $\sqrt{3}$ on warming, and at 298 K, the cell is pseudo-hexagonal. This explains previous anomalous reports of hexagonal symmetry by laboratory X-ray single crystal diffraction, which typically has a lower angular resolution than the powder instruments used here. By fitting a power law,  a transition temperature to a metrically hexagonal cell was estimated to occur at $\sim$335 K. The structure of this phase was investigated using data collected on E9, well above this estimate, at 510 K. This data set does not show any peak splitting that would suggest orthorhombic symmetry (Fig. 2b) and was indexed with a hexagonal cell with \textit{\textbf{a}} = 9.98 \AA \ and \textit{\textbf{c}} = 13.56 \AA.
 \begin{table}
\caption{\label{tab:table1}Refined atomic coordinates \pb \ at 510 K from Rietveld refinement of powder neutron diffraction data. Displacement parameters were constrained to be equal for each species and were Pb:0.019(2), Mn:0.005(3) and O:0.027(4) \AA$^{2}$. The refined lattice parameters were \textit{\textbf{a}}=9.9839(3) and \textit{\textbf{c}}=13.5620(4) \AA.}
\begin{ruledtabular}
\begin{tabular}{cccc}
 $Atom$ &$x$ &$y$ &$z$\\
\hline
 Pb(1) &  0.6122(7)  &   0.6122(7)   &  3/4\\
  Pb(2) &     0.2642(8) &    0.2642(8)   &  3/4\\
    Mn(1)&      0.8331(9)  &   0.1669(9)  &   3/4\\
  Mn(2)&      1/3  &2/3  &0.1445(13) \\
  Mn(3)&      1/2 &1/2&  1/2\\
  Mn(4)&      0 & 0& 0 \\
  O(1)&       0.4875(7)   &  0.3313(7)    & 0.0779(4)   \\
  O(2)&       0.5231(11)  &  0.1727(10)  &  1/4\\
  O(3)&       0.8344(7)   &  0.8344(7) &    0.9256(6)  \\
  O(4)&       0.6689(8)   &  0.6689(8)    & 0.0758(8)  \\
     \end{tabular}
\end{ruledtabular}
\end{table}
  \begin{figure}[tb!]
\begin{center}
\includegraphics[scale=0.3]{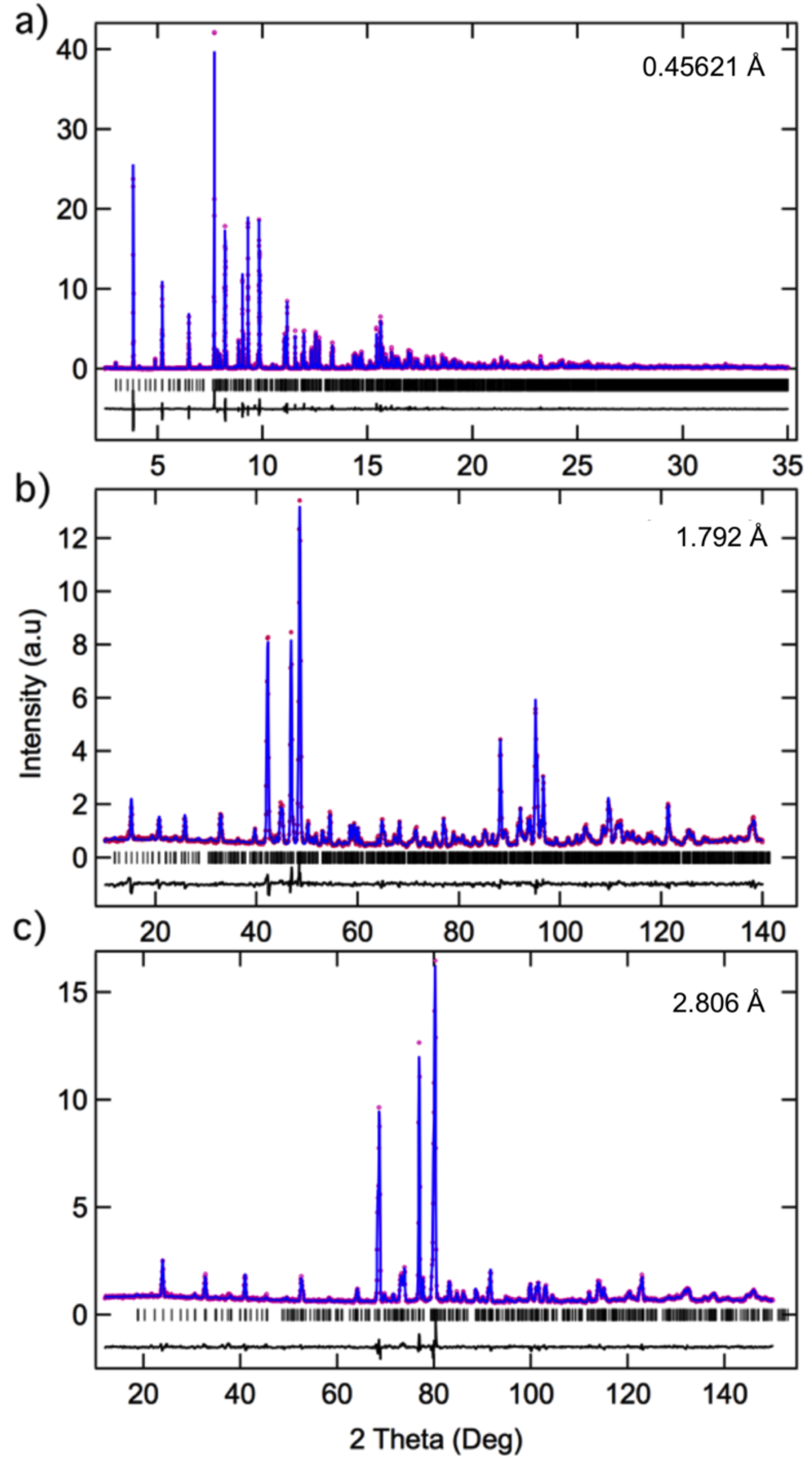}
\caption{(color online) Observed, calculated and difference profiles for the combined Rietveld fit of the $Pnma$ structure of \pb to the room temperature data. The data sets are a) the high resolution synchrotron X-ray and b,c) the neutron powder diffraction  profiles of \pb. The position of reflections are given by the vertical bars.}
\label{Fig1}
\end{center}
\end{figure}
\begin{figure}[tb!]
\begin{center}
\includegraphics[scale=0.18]{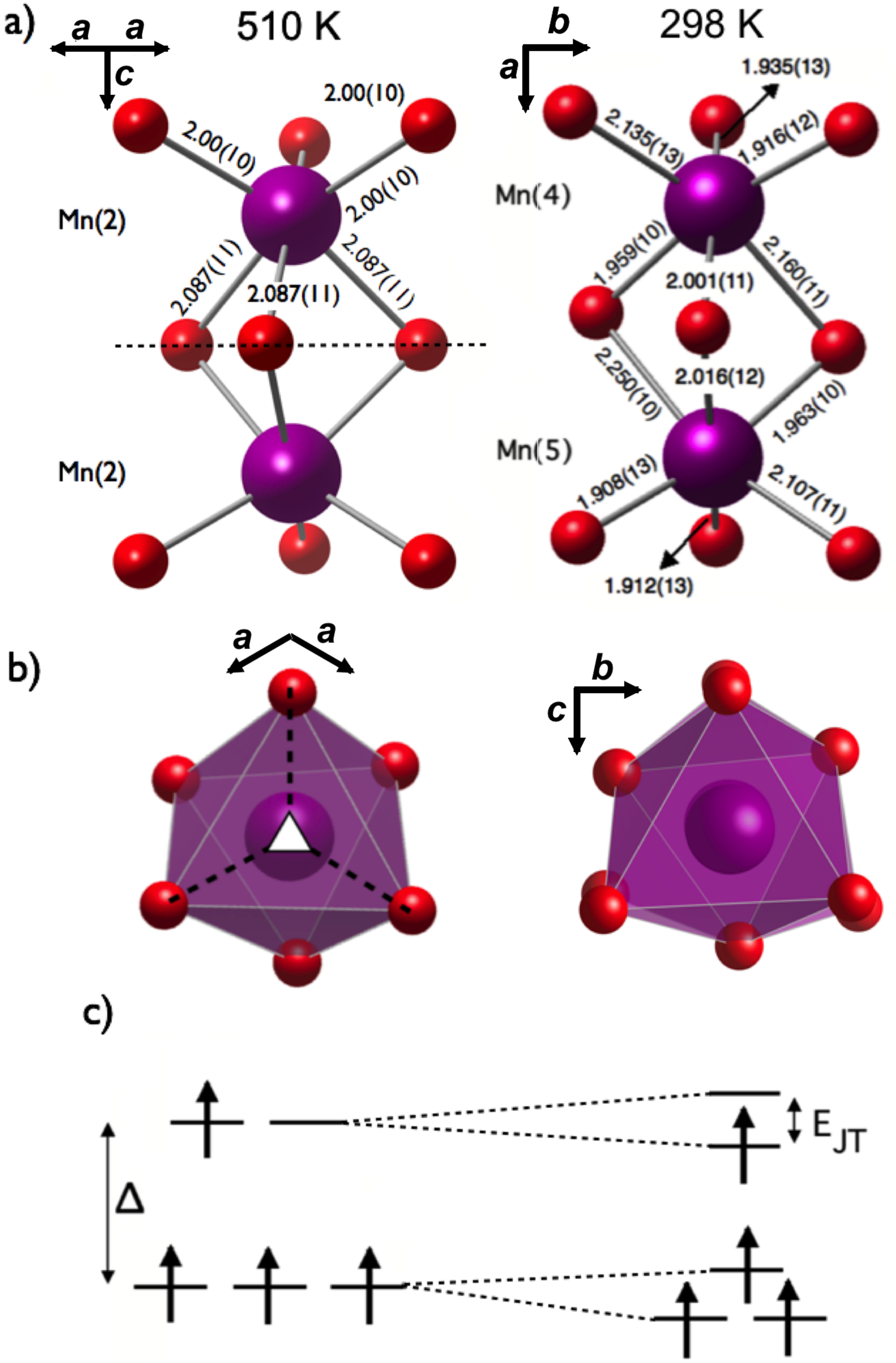}
\caption{(color online) a) Local structure around the interlayer dimer sites in the 510 K \hex \ (left) and 298 K \orth \ (right) phases of \pb. The mirror plane present at 510 K is shown by a dashed line; b) View down [001$_{hex}$] of the interlayer dimer sites highlighting the three-fold rotation axis found in the 510 K structure and its absence at 298 K; c) Schematic of the crystal field splitting of the $e_{g}$ and $t_{2g}$ $d$-orbitals found for a Mn$^{3+}$ cation in both cases.}
\label{Fig1}
\end{center}
\end{figure}
\begin{table*}
\caption{\label{tab:table2}Refined atomic coordinates and displacement parameters for \pb \ at 298 K from combined Rietveld refinement of powder neutron and synchrotron X-ray diffraction data. The displacement parameters were constrained to be equal for the oxygen positions. The refined lattice parameters were \textit{\textbf{a}}= 13.54163(3), \textit{\textbf{b}}=17.22371(4) and \textit{\textbf{c}}=9.99437(2) \AA. }
\begin{ruledtabular}
\begin{tabular}{ccccc}
 $Atom$ &$x$ &$y$ &$z$ &$U_{iso}$\AA$^{2}$\\
\hline
 PB(1)      &0.24957(10)  & 0.44377(5)   & 0.43936(9)   &0.0163(22)  \\
  PB(2)     & 0.25743(8)    &0.11848(5)   & 0.11077(8)  &0.00898(22)     \\
  PB(3)     & 0.77115(10)  & 0.25&  0.98986(12) &0.0736(35) \\
  PB(4)&      0.75915(11)  & 0.25&  0.64518(12)  &0.0908(34)  \\
  Mn(1)&      0.00393(35)  & 0.08244(19)   &0.2444(6) &0.0043(9)       \\
  Mn(2)&      0.00820(25)  & 0.33341(23)   &0.4980(5)   &0.0013(10)    \\
  Mn(3)&      0.51019(25)  & 0.16533(24)   &0.5108(5) &0.0024(10)       \\
  Mn(4)&      0.15265(26)  & 0.41664(28)   &0.7415(5)     &0.0041(11)   \\
  Mn(5)&        0.35704(25)  & 0.08242(26)   &0.7479(5)  &0.001(1)     \\
  Mn(6)&       0  &0  &0&0.0038(14)   \\
  Mn(7)&       0  &0  &0.5&0.0024(14)   \\
  Mn(8)&       0.5099(5)   &  0.25 & 0.7586(8)   &0.0041(13)  \\
  Mn(9)&     0.5092(4)    & 0.25 & 0.2545(9) &0.0045(11)     \\
  O(1)&      0.0786(8)    & 0.4965(7)     &0.3310(13)  &0.0075(4)    \\
  O(2)&       0.4224(8)    & 0.0083(7)     &0.3431(12)  &0.0075(4)    \\
  O(3)&       0.2605(8)    & 0.5083(5)    & 0.1530(7)     &0.0075(4)    \\
  O(4)&       0.9344(8)    & 0.3279(6)     &0.6542(10)    &0.0075(4)   \\
  O(5)&       0.5916(8)     &0.1706(6)     &0.6731(11)    &0.0075(4)    \\
  O(6)&       0.9404(11)    &0.25& 0.4150(15)  &0.0075(4)   \\
  O(7)&       0.5815(12)    &0.25  &0.4131(16)  &0.0075(4)    \\
  O(8)&       0.4421(12)   & 0.25  &0.5945(15)  &0.0075(4)    \\
  O(9)&       0.0884(11)    &0.25  &0.5737(15)   &0.0075(4)  \\
  O(10)&       0.9144(9)    & 0.5854(7)     &0.4294(12)     &0.0075(4)  \\
  O(11)&       0.5766(9)    & 0.9154(7)    & 0.4209(12)   &0.0075(4)    \\
  O(12)&       0.0781(7)     &0.3310(7)    & 0.3284(13)  &0.0075(4)      \\
  O(13)&       0.4329(7)    & 0.1664(7)    & 0.3333(13)  &0.0075(4)    \\
  O(14)&       0.2326(8)    & 0.6625(5)    & 0.1788(8)   &0.0075(4)    \\
  O(15)&       0.2485(10)   & 0.9239(4)    & 0.3898(7)     &0.0075(4)    \\
  O(16)&       0.9337(8)    & 0.4142(6)     &0.4155(12)    &0.0075(4)   \\
  O(17)&       0.5791(8)     &0.0809(6)     &0.4309(12)    &0.0075(4)    \\
    \end{tabular}
\end{ruledtabular}
\end{table*}
 \begin{table*}
\caption{\label{tab:table1}Manganese site splittings, distortions (x10$^{-2}$), and bond valence sum comparison for the 510 K \textit{P6$_{3}$/mcm} and 298 K $Pnma$ phases of \pb.}
\begin{ruledtabular} 
\begin{tabular}{cccccccc}
 $\textit{P6$_{3}$/mcm}$ &  $Site$ & $Distortion$ & $BVS$ &$Pnma$ &  $Site$ &$Distortion$ &$BVS$\\
\hline
 Mn(1) & $12i$ & 1.26& 3.84 &   Mn(1) &$8d$  &2.24&  3.63\\
            & &    &    &    Mn(2)  & $8d$&  1.4&4.00\\
             & &    &    &    Mn(3) &$8d$ &  2.24& 3.45\\
  Mn(2)& $8h$& 2.12& 2.9  &     Mn(4) & $8d$& 4.3&  3.06 \\
         &  &     &   &     Mn(5)  & $8d$&  5.02&3.04\\
  Mn(3)& $6f$&0.69& 3.74  & Mn(6)& $4a$&1.5&     3.95\\
           & &    &    &    Mn(7)& $4b$ & 1.68& 3.35\\
           &  &   &     &    Mn(8)&$4c$ & 1.89&   3.79\\
  Mn(4) & $2b$& 0&3.9  &    Mn(9)& $4c$ & 1.8&  4.05\\
         \end{tabular}
\end{ruledtabular}
\end{table*}
\begin{figure*}[tb!]
\begin{center}
\includegraphics[scale=0.45]{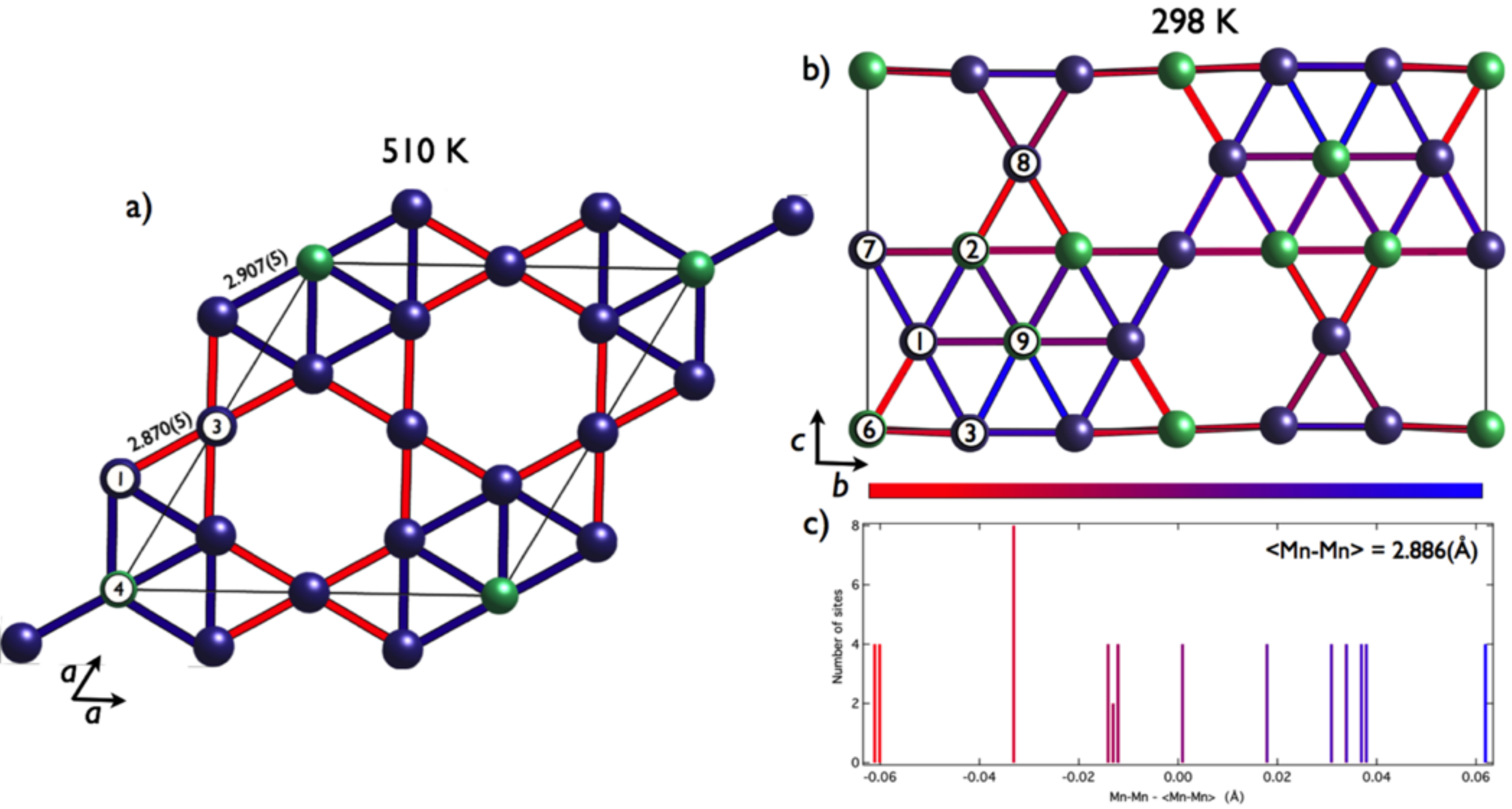}
\caption{(color online) a) Network of Mn-Mn interlayer distances found in the \hex \ phase of \pb at 510 K. The three crystallographically independent Mn sites are indicated according to the results shown in Table III. In both panels sites unambiguously identified as Mn$^{4+}$ by bond valence sums are light green. Sites of intermediate valence are dark blue.; b) Network of Mn-Mn interlayer distances found in the 298 K \orth \ phase of \pb. Note that the axes are related to the \hex \ phase by $\sqrt{3}$.\textit{\textbf{a}} x \textit{\textbf{a}} x \textit{\textbf{c}}. The seven distinct Mn sites are indicated as before. Here the colour coded bonds represent the variation from the average Mn-Mn distance of 2.886 \AA, as shown in the histogram in panel c). }
\label{Fig1}
\end{center}
\end{figure*}
A Rietveld fit of the hexagonal \hex \ structure proposed for Pb$_{3}$Rh$_{7}$O$_{15}$ \ was performed, and rapidly converged with $\chi^{2}$ = 2.59, R$_{wp}$ = 0.063 and R$_{p}$ = 0.048.  No evidence for impurities, or oxygen deficiency was seen. The observed, calculated and difference profiles are shown in Fig. 2c.  The refined cell parameters, atomic coordinates and displacement parameters are shown in Table I. These results are similar to those reported \cite{warm} \ very recently by Volkov $et \ al$ for \pb \ at 573 K. However, comparison of their coordinates (derived from XRD) to the neutron data gave a markedly worse fit after refinement of cell parameters, peak shape and displacement parameters ($\chi^{2}$ = 3.26). The main differences between the two models were found to be small changes in the oxygen parameters due to the greater sensitivity of  neutron diffraction to light atoms.\\
At room temperature, the very high resolution synchrotron powder diffraction confirmed that the unit cell is orthorhombic, and this distortion was also clearly resolved in both neutron powder diffraction data sets (Fig. 3).The systematic absences confirm that the space group is $Pnma$ or $Pn2_{1}a$. The reflections in the class ($k+l$) = odd, which rule out A-centring are extremely weak, of the order of 0.5 \% of the largest reflection in the X-ray data, but significantly stronger in the neutron powder diffraction profiles. This again indicates the importance of oxygen displacements. A trial Le Bail refinement was performed of the profile variables, zero shifts and neutron wavelengths then intensities were calculated using the model of Rasch $et \ al$. With only the scale factors, peak shapes and background functions varied, this refinement converged with $\chi^{2}$ = 3.74. The residuals for the X-ray histogram were R$_{wp}$ = 9.43 and R$_{p}$ = 7.44 \%. For the 2.8 \AA \ neutron histogram, R$_{wp}$ = 7.68 and R$_{p}$ = 5.6 \%. On varying the atomic positions and displacement factors for each atomic species, the refinement converged with $\chi^{2}$ = 3.1 and the residuals improved to R$_{wp}$ = 8.89, R$_{p}$ = 6.88 for the X-ray histogram and R$_{wp}$ = 5.0 and R$_{p}$ = 3.8 for the 2.8 \AA \ neutron histogram. The excellent quality of the fit makes non-centrosymmetric distortions to a \textit{Pn2$_{1}$a} structure unlikely. The major changes are shifts of between 0.1 and 0.15 \AA \ for five of the oxygen positions, as detailed in Table II. The final refined model allows the proposition of charge and orbital order as discussed below.

\section{\label{sec:level1}DISCUSSION}
Previous attempts to estimate the degree of charge separation in \pb \ have suffered from an insensitivity to oxygen displacements or averaging due to twinning in single crystal experiments. In contrast, the combined X-ray and neutron powder diffraction refinement reported here suffers from neither problem and precise atomic positions can be extracted. This is important as both charge and orbital order are evidenced by small shifts of anions in the first coordination sphere. Bond valence calculations \cite{bvs} \ were performed using the experimental coordinates as follows. The individual bond valences ($V_{i}$) are calculated using the observed bond lengths ($R_{i}$):\\
\begin{equation*}
V_{i}=exp(R_{0}-R_{i}/b)
 \end{equation*}\\
 Here $R_{0}$ \ is a tabulated value specific to the type of bond and the metal oxidation state, and $b$ \ is an empirical constant of value 0.37 \AA. The site valence is the sum of the individual bond valences in the first coordination sphere. To ensure an unbiased result, bond valences for each site were calculated using  $R_{0}$ \ values for both Mn$^{3+}$ \ and Mn$^{4+}$ \ cations, then averaged. This procedure was performed for both phases, yielding the results given in Table III, which also shows the relationship between the inequivalent sites in both structures. \\
Turning first to the high temperature $P6_{3}/mcm$ phase, the single crystallographic Mn site in the ribbons between the planes, Mn(2), shows a bond valence sum of 2.9. Furthermore, Mn(4), which is the site that occupies 1/3 of the holes in the Kagom\'e layers, also shows a well defined bond valence sum of 3.9. The only sites that are mixed valence are those which make up the Kagom\'e lattice itself, lying in the range 3.74 - 3.84. These results show that the hexagonal phase of \pb \ should be described as only trivially charge ordered, as the sites which show a significant departure from the expected average bond valence sum value are in very chemically differerent environments. This is similar to the situation found in the multiferroic $RE$Mn$_{2}$O$_{5}$ \ materials, which also have layers of Mn$^{4+}$ cations linked by Mn$^{3+}$ sites. \\
 Much of the interesting physics associated with mixed valence manganite materials, occurs due to cooperative orbital ordering of degenerate Mn$^{3+}$ \ $e_{g}^{1}$ electrons. In order to search for possible Jahn-Teller distortions in both phases,  a distortion parameter \cite{dist} \ was calculated for each site as follows:\\
\begin{equation*}
D=\frac{1}{n}\sum_{i=1}^{n}\frac{|l_{i}-l_{av}|}{l_{av}}
 \end{equation*}\\
 Here l$_{i}$ are the individual Mn-O bond lengths and l$_{av}$ is the average Mn-O bond length in each octahedron. All of the manganese sites were found to be highly symmetric at 510 K, with distortion parameters falling between 0 and 2.12x10$^{-2}$. This is especially notable for the interlayer Mn$^{3+}$ dimer, which is highly regular due to the presence of a mirror plane and three fold site symmetry in the $P6_{3}/mcm$ space group (Figs. 4a and 4b). Finally, the refined Mn-Mn distances in the 1/3 filled Kagom\'e layers are shown in Fig. 5a. This distance is found to be slightly larger around Mn(4) (2.907(5) \AA), which is as expected due to the higher formal charge inferred from the bond valence sums for this site. The Mn-Mn distance in the rest of the layer is slightly shorter (2.870(5) \AA) and highly regular by symmetry, which implies frustration of spin, charge and orbital degrees of freedom as discussed in more detail below.\\
Turning now to the room temperature structure, and in particular the splitting of manganese sites shown in Table III, it is instructive to begin with those sites which change least. The site at the centre of the 'stars' in the layers, Mn(4), does not split. Furthermore, the extracted bond valence sum for the equivalent site in the $Pnma$ structure, Mn(9), shows that no charge re-distribution has occurred (4.05). The site distortion (which was zero by symmetry in the high temperature structure) also remains small (1.8x10$^{-2}$). A similar picture, as regards charge degrees of freedom, is found for the Mn sites which occupy the interlayer dimers. These are split into two sites (Mn(4) and Mn(5)), both of which show bond valence sums very close to 3+ (3.06 and 3.04 respectively). However, these sites show a dramatic increase in site distortion (4.3 and 5.02x10$^{-2}$), which is evidence for a lifting of $e_{g}^{1}$ \ orbital degeneracy. This is supported by the refined Mn-O bond distances shown in Figs. 4b and 4b. Both sites show an elongation imposed upon the intrinsic trigonal distortion\cite{banaruo}. For example, Mn(4) has two long Mn-O  bonds (2.135(13) and 2.160(11) \AA) and a group of four shorter bonds in the range 1.935(13) - 2.001(11) \AA. The corresponding change in the energy of the $d$-orbitals is shown schematically in Fig. 4c, and can be seen to be similar to that seen in classic orbital ordering examples\cite{goodenough} such as LaMnO$_{3}$, where the \textit{d}z$^{2}$ \ orbital is preferentially occupied. This electronically driven effect breaks the three fold symmetry found on these sites and lowers the crystallographic symmetry to orthorhombic. \\
Subtle metal displacements are also found in the $Pnma$ structure, as shown in Fig. 5b. This is reminiscent of other spin or charge frustrated  systems which undergo long-range order. These distortions help to select a single groundstate configuration out of the degenerate manifold of states. In spinel structured materials, such as CuIr$_{2}$S$_{4}$ or Fe$_{3}$O$_{4}$ \ for example, metal positions shift such that octamers\cite{rad} \ or trimers\cite{senn} \ are found. In the present case, the most clearly seen effect is an expansion around the Mn$^{4+}$ sites which occupy the holes in the Kagom\'e lattice. As these sites do not split, or change in oxidation state upon cooling to room temperature, this may indicate a decoupling of the mixed-valence sites which make up the Kagom\'e network (Fig. 1a).
This interpretation is in accordance with the results shown in Table III, as these two sites are found to split into a total of six in the \orth \ phase. For each site, a splitting into one site of close to Mn$^{4+}$ valence is found (Mn(2)) and Mn(6)) and one with a valence closer to  Mn$^{3+}$ (Mn(3) and Mn(7)). The third site in each case shows a bond valence almost unchanged from the high temperature value. The Kagom\'e layers are thus segregated into three equally populate types of sites as shown in Fig. 6. Notably, none of these sites shows an increased distortion, showing that orbital degrees of freedom are still frustrated at room temperature.  The charge order over these sites explains why A-centring is lost in the room temperature structure as the sites at (000) and (0,1/2,1/2) are no longer crystallographically related. \\
The three-fold charge order found here (Fig. 6) is different to that seen in materials with frustrated triangular lattices\cite{ag,lu} \ such as AgNiO$_{2}$ or LuFe$_{2}$O$_{4}$. These materials show a 2:1 ratio of  charge rich and poor sites, which may support a (anti)-ferroelectric polarisation\cite{comul}. Although spin models showing three-sublattice order have been widely studied on the Kagom\'e lattice\cite{kag}, the arrangement found here does not match any of the commonly found groundstates such as the Q=0 or $\sqrt{3}$ x $\sqrt{3}$ \ structures. This may be an indication of the importance of longer range interactions which these models do not capture, or might indicate a role for the remaining frustrated orbital degrees of freedom or the Mn$^{4+}$ sites which partially occupy the voids. \\
\begin{figure}[tb!]
\begin{center}
\includegraphics[scale=0.32]{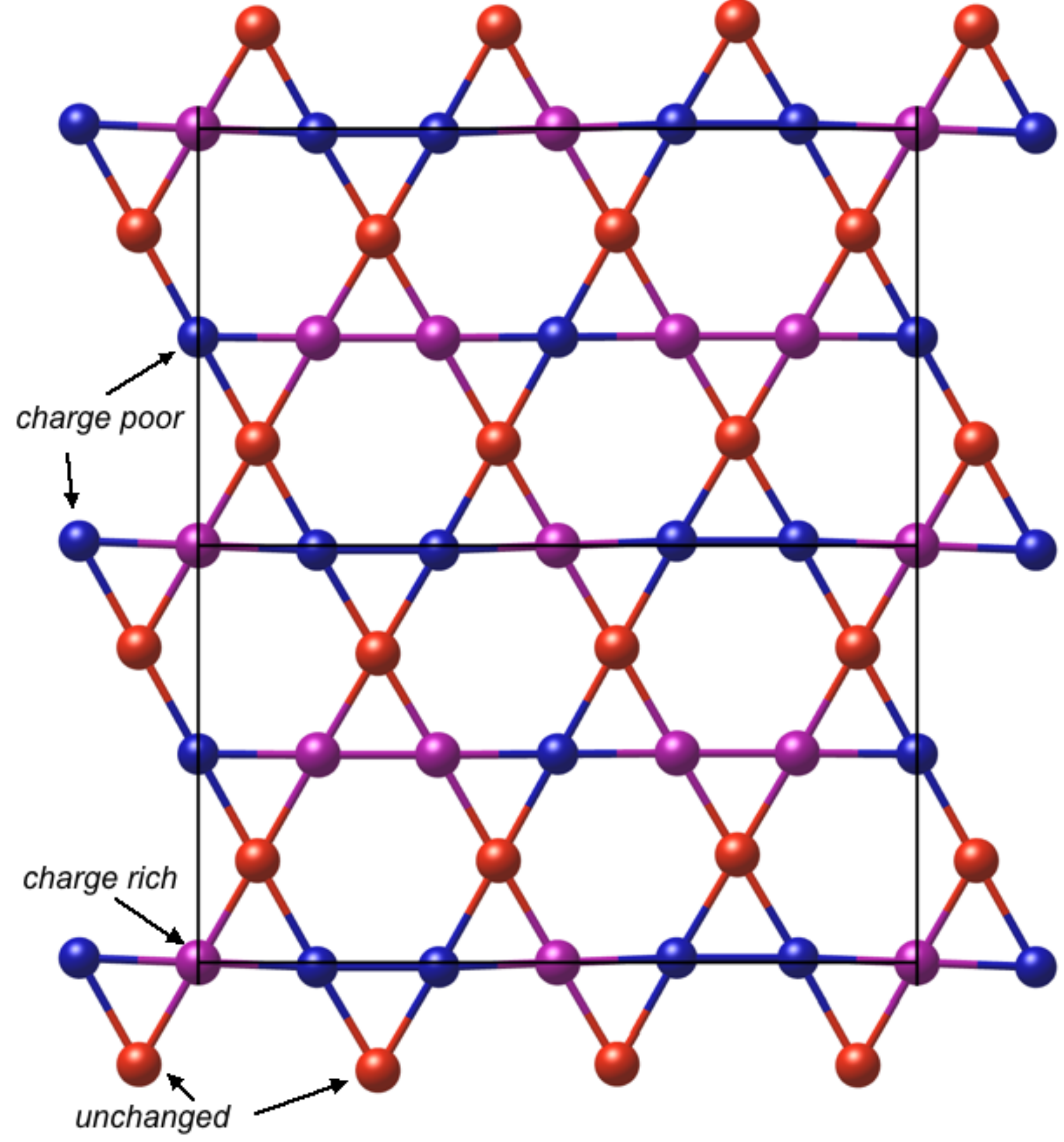}
\caption{(color online) Three-fold charge ordering on the sites which make up the Kagom\'e layers in the $Pnma$ \ structure at room temperature. Sites which are charge rich are defined as those with bond valence sums $>$ 3.9 and sites which are charge poor are defined as those which have BVS's $<$ 3.45. Unchanged sites have BVS's in the range 3.63 to 3.79.}
\label{Fig1}
\end{center}
\end{figure}
Finally, the partial charge and orbital ordering determined in this work for \pb \ at room temperature has implications for the physical properties, in particular the dielectric constant measurements reported elsewhere. The charge ordering pattern does not support either a ferroelectric or anti-ferroelectric polarisation. The peak in dielectric constant\cite{dielectric} \ which appears around 150 K is therefore probably related to some as yet unknown structural rearrangement which occurs below room temperature. This is in keeping with reports of a small specific heat anomaly\cite{hcap} \ and a change in slope of resistivity  around 250 K. Further detailed measurements  using a combination of X-ray and neutron diffraction measurements, and investigations of the spin order at low temperatures will help to shed more light on this postulated transition.\\
 \section{\label{sec:level1}CONCLUSIONS}
In summary, the above results resolve a long standing structural question, the origins of the extremely complex distortions found in \pb \ at room temperature. In sharp contrast to previous studies\cite{str1,str2,str3,rasch,warm,mag1}, bond valence sums and distortions of the local coordination environment around the Mn cations show a charge and orbital ordering occurs between 510 and 298 K. However, several sites are trivially charge ordered at all temperatures, and the only sites which split in the edge sharing layers are those which make up a Kagom\'e network.\\

This work was partially performed during an EPSRC studentship at the University of Edinburgh, and in the Novel Materials group at the Helmholtz-Zentrum Berlin. J.P. Attfield and D.N. Argyriou are thanked respectively for training, advice and encouragement. A.N. Fitch, J.W.G. Bos and P.F. Henry  are thanked for assistance with the diffraction measurements. ESRF and HZB are acknowledged for the award of beam time.

\end{document}